\documentclass[superscriptaddress,twocolumn,floatfx,amsmath,amssymb,prl]{revtex4-2}
\usepackage{amsmath,amssymb}
\usepackage{graphics}
\usepackage{epsfig}
\usepackage{extarrows}
\usepackage{MnSymbol}
\usepackage{mathrsfs}
\usepackage[usenames]{color}
\usepackage{appendix}

\begin{document}

%\title{Measuring loss constants of ultracold atoms via Lellouch-L\"uscher type relation}
\title{Lellouch-L\"uscher relation for ultracold few-atom systems under confinement}
\author{Jing-Lun Li}
\email{Jinglun.Li@ist.ac.at}	
\affiliation{Institut f\"{u}r Quantenmaterie and Center for Integrated Quantum Science and
	Technology IQ$^{ST}$, Universit\"{a}t Ulm, 89069 Ulm, Germany}
\affiliation{Institute of Science and Technology Austria (ISTA), 
Am Campus 1, 3400 Klosterneuburg, Austria
}%    
\author{Paul S. Julienne}
\affiliation{Joint Quantum Institute, University of Maryland, and the National
Institute of Standards and Technology (NIST), College Park, MD 20742, USA}
\author{Johannes Hecker Denschlag}
\affiliation{Institut f\"{u}r Quantenmaterie and Center for Integrated Quantum Science and
	Technology IQ$^{ST}$, Universit\"{a}t Ulm, 89069 Ulm, Germany}
\author{Jos\'{e} P. D'Incao}
\email{Jose.DIncao@umb.edu}
%\affiliation{JILA, NIST, and the Department of Physics,
%University of Colorado, Boulder, CO 80309, USA}	
\affiliation{Department of Physics, University of Massachusetts Boston, Boston, MA 02125, USA}	
\date{\today}

\begin{abstract}
We derive an analog of the Lellouch-L\"uscher (LL) relation for few-body bosonic systems, linking few-body scattering loss rates to the energies and widths of the corresponding harmonically trapped few-body states. Three-body numerical simulations show that the LL relation applies across a broad range of interaction strengths and energies and allows the determination of scattering rates within a single partial wave. Our work establishes a robust theoretical framework for understanding the role of the finite-volume effect in few-body observables in optical lattice and tweezer experiments, enabling precise determination of multi-body scattering rates.
\end{abstract}
\maketitle

Over the past few decades, trapped ultracold atoms and molecules have become a versatile platform for exploring and exploiting quantum phenomena, driving advances in atomic, molecular and optical physics, condensed matter, chemistry, and quantum information science and technology \cite{block2005NP,block2005AAMOP,schafer2020NRP,lester2018prl,kaufman2018AAMOP,anderegg2019Sc,krylov2021PCCP,Bloch:2008,Lewenstein:2007}. This success stems from the unparalleled level of control provided by ultracold atomic systems over internal states, interatomic interactions, and trapping dynamics \cite{Chin:2010,kaufman2012prx,fortagh2007rmp,gauthier2021AAMOP,Ruttley:2024}. More recent advances in optical lattices and optical tweezers have further expanded the versatility of ultracold atoms by introducing deterministic control of the atom number (up to a few) in a single trap site \cite{Ott:2016,Kaufman:2021,Andersen:2022,Pampel2024}. 
These techniques enable novel approaches to quantum simulation \cite{brown2019prx} and allow deeper insight into the fundamental processes governing chemical reactions \cite{Liu:2018,Liu:2019,Zhang:2020,Ruttley:2023}. Furthermore, they pave the way for using trapped few-atom systems for quantum information applications, the exploration of multi-body interactions \cite{Goban:2018}, and the study of many-body states created atom by atom \cite{Wenz:2013}.

In many ways, atoms and molecules trapped in optical lattices or tweezers form an ideal system for studying various fundamental aspects of few-body interactions. While in ultracold gas experiments few-body interactions are typically characterized by scattering rates, in trapped systems the key observables are the on-site interatomic interaction and the lifetime of individual trapped states \cite{Wenz:2013,Mies:2000,Greiner:2002, Ospelkaus:2006, Amato-Grill:2019, Will:2010,Goban:2018,mark2020PRR,venu2023NT}. Optical tweezers, in particular, have enabled the creation of fully isolated two- and three-atom systems, allowing measurements of such physical observables at the single-event level \cite{Xu:2015, Hood:2020, Reynolds:2020}. 
While offering unprecedented control over few-body systems, such experiments also encounter two central challenges: how to relate measurements in confined setups to free-space scattering parameters and how confinement itself influences collision dynamics and system stability. Similar challenges arise in high-energy physics, particularly in lattice quantum chromodynamics, where calculations are performed in a finite volume. A breakthrough in that context was the development of the well-known Lellouch-L\"uscher (LL) relations \cite{Lüscher:1986s,Lüscher:1986,Maiani:1990,Lellouch:2001,Beane:2011,Hammer:2017,Müller:2021,Beane:2014,Stellin:2021,Bubna:2024}, which provide a rigorous way to extract scattering observables --such as decay rates and cross sections-- from the discrete energy levels of finite-volume simulations. A closely related treatment was established in hadronic atoms and is commonly referred to as the Deser–Trueman formula \cite{Deser:1954,Trueman:1961,Thomas:1980,Gasser:2008,Curceanu:2019}. 
%\textcolor{blue}{This approach has also been extended to systems with long-range forces relevant for hadronic atoms \cite{Beane:2014,Stellin:2021,Bubna:2024}.}
The development of a similar approach for trapped few-atom systems has the potential to provide a more accurate and controllable procedure to determine few-atom scattering rates than in ultracold bulk gases. 
Isolated few-body systems allow for a clearer and more systematic way to distinguish scattering rates from processes involving varying numbers of atoms \cite{Reynolds:2020}, a task that proves to be more challenging in bulk gases \cite{kraemer2006NT,stecher2009NP,ferlaino2009PRL,zenesini2013NJP}. Moreover, the ability of preparing trapped few-body states in well-defined quantum states \cite{zurn2013prl,Kaufman:2015,Goban:2018,Liu:2019,Spence:2022,brooks:2022,Shaw:2025} can enable the direct measurement of individual partial-wave scattering rates, providing a much more sensitive probe for resonance and interference effects often obscured by thermal effects in gases \cite{dincao2004prl,huang2015PRA}.

In this Letter, we derive the analog of the LL relation for ultracold few-body systems under harmonic confinement and provide numerical evidence for its validity. In particular, we consider a system of three identical bosons in an isotropic harmonic trap and focus on extracting the free-space energy-dependent three-body loss (recombination) rate coefficient, $L_3(E)$, characterizing the process where three free atoms collide at energy $E$ to form a diatomic molecule. 
Our key finding is an LL relation that connects $L_3(E_q)$ at the energy $E_q>0$ ($q=0,1,2, \dots$) of the $q$-th gas-like trapped three-body state in the lowest partial wave \cite{Blume:2012,Sykes:2014,DIncao:2018E} to its width, $\Gamma_q$: 
\begin{equation}
L_3(E_q)=C\frac{\hbar^4 }{m^3}\frac{\Gamma_q}{\omega_{\rm ho}[E_q^2-(\hbar\omega_{\rm ho})^2]} \label{eq:L3}
\end{equation}
 where $m$ is the atomic mass, $\omega_{\rm ho}$ the angular trap frequency and $C=72\sqrt{3}\pi^3$, a universal constant. Conversely, if $L_3(E)$ is known, the LL relation predicts the lifetime, $\hbar/\Gamma_q$, of the trapped three-body states \cite{exL3}. We numerically verify our LL relation through simulations of $^{85}$Rb and $^{87}$Rb atoms using the hyperspherical adiabatic representation \cite{Suno:2002,wang2011pra,Lipaper}. We show that the LL relation remains valid for states in which $|a|\ll r_{\rm ho}{(q)}\simeq({E_q/m \omega_{\rm ho}^2})^{1/2}$, where $r_{\rm ho}{(q)}$ is the characteristic size of the $q$-th trap state and $a$ is the two-body scattering length characterizing the strength of the interatomic interaction \cite{Chin:2010}.  We generalize the LL relation to a system of $N$ identical bosons opening up a path for the determination of multi-body scattering rates \cite{Goban:2018,mehta2009prl}. These results provide a direct theoretical framework for interpreting few-body observables in optical lattices and tweezers experiments, where finite-volume effects play a central role.
 \begin{figure}[t]
 \centering
  \resizebox{0.48\textwidth}{!}{\includegraphics{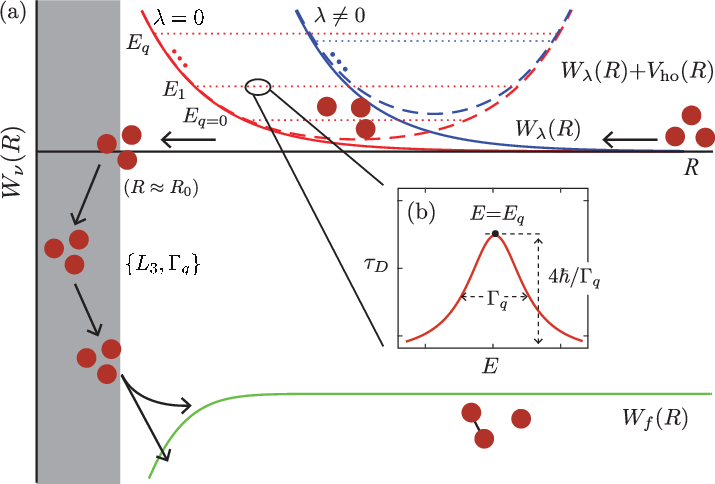} }
 \caption{\label{fig:trap3} 
(a) Schematic illustration of three-body losses in free space and in a harmonic trap. In free space, three atoms in continuum channels $W_{\lambda}(R)$ (solid) undergo inelastic transitions near $R \lesssim R_0$ (shaded) to atom-dimer channels $W_{f}(R)$, characterized by the recombination rate constant $L_3$. In a trap, discrete states $E_q$ (dotted) form in $W_\lambda(R)+V_{\rm ho}(R)$ channels (dashed) and decay into atom-dimer channels with rate $\Gamma_q/\hbar$ \cite{sm}.  
(b) Trapped-state energies $E_q$ and widths $\Gamma_q$ manifest as Lorentzian resonances in the time delay $\tau_D(E)$ (see text).
}
\end{figure}

In adiabatic hyperspherical representation \cite{Suno:2002,wang2011pra}, three-body observables are determined from the solution of the hyperradial Schr\"{o}dinger equation \cite{Suno:2002,wang2011pra}
    \begin{align}
&\left[-\frac{\hbar^2}{2\mu}\frac{d^2}{dR^2}+W_{\nu}(R)-E\right]F_{\nu}(R)\nonumber\\
&~~~~~~~~~~~+\sum_{\nu'\neq\nu}W_{\nu\nu'}(R)F_{\nu'}(R)=0,\label{Schro}
\end{align}
with the adiabatic potentials $W_{\nu}$ and the non-adiabatic couplings $W_{\nu\nu'}$ obtained by solving the adiabatic hyperangular equations at fixed values of the hyperradius $R$ (see Supplemental Material \cite{sm} for more details\nocite{Bolda:2002,Kohler:2005,Jack:2002,Ratzel:2021,sphv}). Here, $\nu$ is the set of quantum numbers necessary to characterize each channel, $F_{\nu}$ is the three-body hyperradial wave function and $\mu=m/\sqrt{3}$ the three-body reduced mass.
In Fig.~\ref{fig:trap3}(a) we show the adiabatic potentials $W_{\nu}$
for three-body loss in free space and in a harmonic trap. 
Here, $W_{\nu=\lambda}$ are continuum channels that describe the collision between three free atoms or, in a trap, support discrete three-body states. 
They are labelled by the hyperangular momentum $\lambda{\ge 0}\in\mathbb{Z}$, 
which sets the hyperradial centrifugal barrier at large $R$ \cite{sm}. (For bosonic systems the lowest value of $\lambda$ is $\lambda_{\rm min}=0$ while for fermionic systems $\lambda_{\rm min}\neq0$ \cite{Dincao:2018}.)
In Fig.~\ref{fig:trap3}(a), $W_{\nu=f}$ denotes an atom–dimer channel, 
which serves as the primary decay pathway leading to atomic losses. 

The central physical quantities in the LL relation (\ref{eq:L3}) can be extracted from the solutions of Eq. (\ref{Schro}) in both free space and trapped systems.
In free space, the rate constant for three-body recombination at collision energy $E$ is defined as
\begin{align}
    L_3(E)=\sum_J\sum_{\lambda,f}96\pi^2\hbar\frac{(2J+1)}{\mu k^4}|S^J_{f\lambda}|^2=\sum_J L_3^J(E),\label{L3Smat}
\end{align}
with $k=(2\mu E/\hbar^2)^{1/2}$, $f$ running over all final $W_f$ (atom-dimer) channels, $\lambda$ running over the initial $W_\lambda$ (three-body continuum) channels, and the corresponding $S$-matrix obtained from the solution of Eq.~(\ref{Schro}) \cite{wang2011pra}. In the above equation, $L_3^{J}$ is the partial-wave contribution to $L_3$ for the total orbital angular momentum, $J$. For ultracold bosonic gases, the total rate $L_3(E)$ is dominated by the $J=0$ partial rate, with all other $J$ contributions vanishing as $E\rightarrow0$ \cite{esry2001pra,dincao2005prl}, only becoming important for energies beyond $E_c=\hbar^2/ma^2$ \cite{dincao2004prl}. In the present work, we study $L_3^{J=0}$ recombination and its connection to the lifetime of the $J=0$ trapped three-body states.  

By introducing the harmonic trap $V_{\rm ho}(R)=\mu \omega_{\rm ho}^{2}R^2/2$ into Eq.~(\ref{Schro}), each $W_\lambda$ channel will support a series of trapped states with energy $E_q$ and width $\Gamma_q$, arising from the decay into $W_f$ channels [illustrated in  Fig. \ref{fig:trap3}(a); see further details in Ref.~\cite{sm}]. 

Here, we focus on trapped states associated with the $\lambda=\lambda_{\rm min}=0$ channel, as this is the dominant channel for $L_3^{J=0}(E)$ \cite{wang2011pra}. As we will see below, our numerical calculations show that the widths $\Gamma_q$ for trapped states associated with $\lambda\ne0$ are extremely narrow, indicating the weak decay from such channels. (Note that we omit $J$ and $\lambda$ in our following notation unless stated otherwise.) 
Here, the energies and widths of the trap states are determined via the time delay $\tau_D(E)$, 
\begin{align}
 \tau_D(E)\equiv 2\hbar\frac{d}{d E}\delta_S(E),\label{tau}
\end{align}
with the eigenphase-shift sum, $\delta_S$, obtained from the corresponding $S$-matrix elements \cite{Nielsen:2002} extracted from the solutions of Eq.~(\ref{Schro}) with the trap $V_{\rm ho}(R)$. As usual, near the energy of a trapped state, the time delay exhibits a Lorentzian line shape [see Fig. \ref{fig:trap3}(b)], reaching its maximum at $E=E_q$, with a width determined by $\Gamma_q=4\hbar/\tau_D(E_q)$. See Ref.~\cite{sm} for details on our numerical procedure to determine $E_q$ and $\Gamma_q$.

After having introduced the theoretical framework for calculating $L_3(E)$, $E_q$ and $\Gamma_q$, we now present a simple analytical approach to derive the relationship between free-space and in-trap observables, i.e., the  LL relation (\ref{eq:L3}). We assume weak interactions ($|a|/a_{\rm ho} \ll 1$, where $a_{\rm ho}=\hbar^{1/2}/(m\omega_{\rm ho})^{1/2}$ is the oscillator length), but also weak trapping ($r_{\rm vdW}/a_{\rm ho}\ll 1$, with $r_{\rm vdW}$ being the van der Waals (vdW) length \cite{Chin:2010,sm}).
Our key observation is that the mechanism controlling both $\Gamma_q$ and $L_3$ is identical, i.e., the inelastic transitions occur primarily at short distances $R \leq R_0 \sim r_{\rm vdW}$, as illustrated in Fig.~\ref{fig:trap3}(a).
Accordingly, we define the following relation \cite{Werner:2006,Petrov:2004w,Nielsen:2002}:
\begin{equation} \label{anal1m}
\frac{\Gamma_q}{\hbar}=\frac{L_3(E_q)}{3}\mathcal{D}_q,
\end{equation} 
where the factor 3 accounts for the loss of three atoms per event, and $\mathcal{D}_q$ is the reactive three-body probability density defined as the probability density to find particles in the $q$-th state within the reaction region ($R \leq R_0$), with hyperspherical volume $\mathcal{V}_0=\pi^3R_0^6/6$. In Ref.~\cite{sm} we show that $\mathcal{D}_q$ is closely related to the three-body correlation function \cite{Reynolds:2020}.
In hyperspherical representation, $\mathcal{D}_q$ is given by
\begin{eqnarray} \label{pqr}
\mathcal{D}_q=\frac{1}{\mathcal{V}_0}\int_0^{R_0}F_q(R)^2dR,
\end{eqnarray}
where $F_q$ is the (non-interacting) hyperradial wavefunction for the $q$-th trapped three-body state \cite{Werner:2006a} associated with the $W_\lambda$ channel. For $\lambda=0$, $F_q$ is given by
\begin{equation} \label{fqnon}
F_{q}(R)=\frac{R^{5/2}}{{N^{1/2}_qa_{\rm ho}^3}}L_q^{(2)}\left(\frac{R^2}{\sqrt{3}a_{\rm ho}^2}\right)\exp\left(-\frac{R^2}{2\sqrt{3}a_{\rm ho}^2}\right),
\end{equation}
 with energy $E_{q}$$=$$(2q+3)\hbar\omega_{\rm ho}$. Here, $L_q^{(\alpha)}$ is the generalized Laguerre polynomial and $N_q=3\sqrt{3}(q+1)(q+2)/2$ is a normalization constant. To leading order, the reactive probability density can be expressed as \cite{sm},
\begin{equation}
\mathcal{D}_{q} = \frac{m^3\omega_{\rm ho}(E_q^2-\hbar^2 \omega_{\rm ho}^2)}{24 \pi^3\sqrt{3}\hbar^5}, \label{Pq}
\end{equation}
independent of the specific choice of $R_0$. Substituting Eq.~(\ref{Pq}) into (\ref{anal1m}), we finally obtain the LL relation (\ref{eq:L3}), 
\begin{equation}
\Gamma_q=\frac{1}{C}\frac{m^3}{\hbar^4} L_3(E_q)[E_q^2-(\hbar\omega_{\rm ho})^2] \omega_{\rm ho}, \label{gammarM}
\end{equation}
with $C=72\sqrt{3}\pi^3$. The above derivation of the bosonic $J=0$ ($\lambda_{\rm min} = 0$) LL relation [Eqs.~(\ref{eq:L3}) and (\ref{gammarM})] can be readily extended to other bosonic and fermionic systems with arbitrary $J$ ($\lambda_{\rm min}$) \cite{Dincao:2018}.

We note that beyond the weakly interacting regime 
($|a|/a_{\rm ho} \ll 1$), interactions induce measurable shifts in both 
the energy and the spatial extent of the trap states. 
Since $\mathcal{D}_q$ [Eq.~(\ref{Pq})] quantifies the probability density within the 
reaction region, it becomes sensitive to interaction-induced changes of the state size. 
The use of the numerically obtained energy $E_q$ in $\mathcal{D}_q$ [Eq.~(\ref{Pq})] and 
the LL relations [Eqs.~(\ref{eq:L3}) and (\ref{gammarM})], rather than the noninteracting 
form $E_q=(2q+3)\hbar\omega_{\rm ho}$, incorporates leading-order interaction-induced
modifications to the reaction volume~\cite{Bolda:2002,Kohler:2005} while preserving 
the model structure. Furthermore, as $q$ increases, the size of the trapped state grows 
as $r_{\rm ho} (q)\simeq(2q+3)^{1/2}a_{\rm ho}$, reducing interaction effects and 
improving the accuracy of the LL relation. [Accordingly, the condition $|a|/r_{\rm ho}(q)\ll1$ 
provides a more appropriate criterion for its validity than $|a|/a_{\rm ho}\ll1$.] 
These aspects were verified numerically and discussed 
in more depth in Ref.~\cite{sm}.

We now test the validity of the LL relation through three-body numerical simulations of $^{85}$Rb and $^{87}$Rb atoms. Our numerical model employs reduced interatomic interaction potentials derived from Born-Oppenheimer potentials while incorporating the full atomic hyperfine spin structure \cite{Lipaper,Lipapertwo} (see also Ref.~\cite{sm}). This interaction model encompasses more than 100 atom-dimer channels, with three-body potentials that span an energy range several thousands times the characteristic vdW energy scale, $E_{\rm vdW}=\hbar^2/mr_{\rm vdW}^2$, below the three-body breakup threshold ($E\leq0$) \cite{sm}. 
%\section{Numerical results }
 \begin{figure}[t]
 \centering
  \resizebox{0.48\textwidth}{!}{\includegraphics{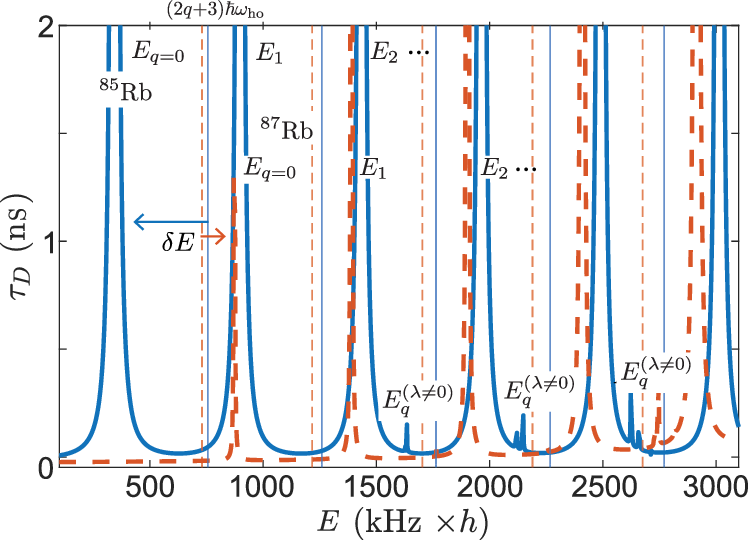} }
 \caption{\label{fig:Tdelay} A typical time delay plot for $^{85}$Rb at $\omega_{\rm ho}$=$2\pi\times$252 kHz or, equivalently, $|a|/a_{\rm ho}$=1.11 (solid line) and for $^{87}$Rb at $\omega_{\rm ho}$=$2\pi\times$243 kHz, or $|a|/a_{\rm ho}$=0.24 (dashed line). Both frequencies correspond to $a_{\rm ho}$=$5 r_{\rm vdW}$. The vertical lines indicate the energy levels of three non-interacting atoms, $(2q+3)\hbar\omega_{\rm ho}$. Arrows indicate the shift $\delta E\propto a/a_{\rm ho}$ from the ground state energy $3\hbar\omega_{\rm ho}$ due to interactions. $E_{q}$ denotes the energies for states of the $\lambda=0$ channel and $E_{q}^{(\lambda \neq 0)}$ otherwise.}
\end{figure}

 \begin{figure*}[t]
 \centering
  \resizebox{1.\textwidth}{!}{\includegraphics{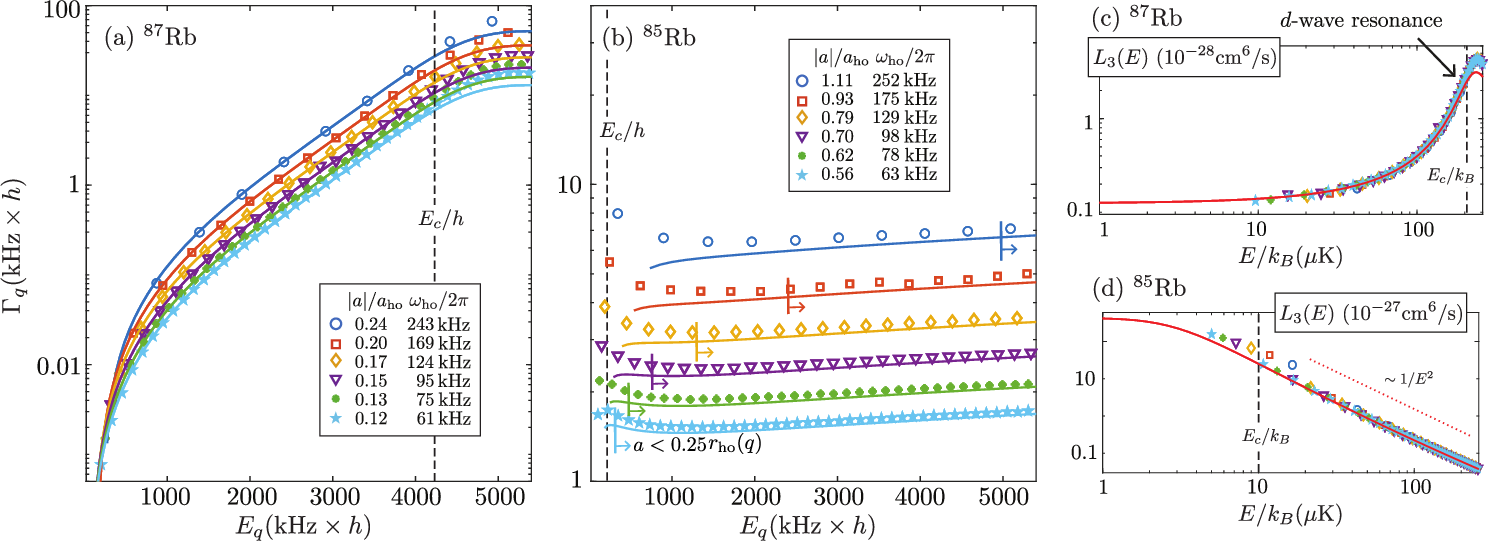} }
\caption{\label{fig:ER} Numerical values for the width $\Gamma_q$ of the three-body states (symbols) as a function of their energy $E_q$ for (a) $^{87}$Rb, and (b) for $^{85}$Rb. Solid lines correspond to the results for $\Gamma_q$ from the LL relation (\ref{gammarM}), obtained from our free-space numerical simulations of $L_3(E)$. { The colored arrows in (b) mark the regime $a<0.25 r_{\rm ho}(q)$, and all data points in (a) fall within this regime.} {For both Rb isotopes the frequencies correspond to $a_{\rm ho}/r_{\rm vdW}$ = 5, 6, 7, 8, 9 and 10.} Symbols in (c) and (d) correspond to $L_3(E_q)$ from the LL relation (\ref{eq:L3}) obtained from the numerical calculations for $E_q$ and $\Gamma_q$ shown in (a) and (b). The vertical dashed lines indicate the value of $E_c$=$\hbar^2/(ma^2)$.}
\end{figure*}
The two isotopes of Rb atoms we use have vastly different physical properties: at zero-magnetic-field, the effective interaction between $^{85}$Rb atoms is strong and attractive, characterized by the large and negative value of $a=-5.56$ $r_{\rm vdW}$ while for $^{87}$Rb the effective interaction is weak and repulsive with $a=1.21$ $r_{\rm vdW}$ \cite{sm}. In the trap, interactions induces energy shifts ($\delta E \propto a/a_{\rm ho}$) from the non-interacting energy levels, $(2q+3)\hbar\omega_{\rm ho}$ \cite{Blume:2012,sm}.
This effect can be observed in Fig. \ref{fig:Tdelay} for the time delay for three $^{85}$Rb (solid line) and $^{87}$Rb (dashed line) atoms under strong confinement ($a_{\rm ho}=$5 $r_{\rm vdW}$). Figure~\ref{fig:Tdelay} shows $\lambda=0$ trapped states for both atomic species manifested by the resonance feature (sudden increase) of the time delay near $E=E_q$, shifted by $\delta E$ from the non-interacting energies (vertical lines). Note that the narrow and unresolved features in Fig.~\ref{fig:Tdelay} are associated with three-body trap states belonging to $\lambda\neq0$ channels, highlighting their relative insignificance compared to the $\lambda=0$ states \cite{wang2011pra}. 
More notably, while the resonances for $^{85}$Rb are generally broad, with $\Gamma_q$ largely insensitive to $E_q$, those for $^{87}$Rb display a clear transition from narrow to broad as $E_q$ increases.
In the following, we show that the LL relation provides a clear interpretation of these two distinct physical behaviors, attributed to the energy dependence of $L_3(E)$.

In Fig.~\ref{fig:ER}, we validate our LL relation (\ref{gammarM}) by comparing the calculated $\Gamma_q$ values (symbols) with LL predictions (solid lines), using our results for $L_3(E)$.
Figure~\ref{fig:ER} presents the results for trapping frequencies ranging from 60 kHz to 250 kHz, showing excellent agreement with the LL relation~(\ref{gammarM}) in the regime $|a|/a_{\rm ho}\ll1$, with deviations emerging as $|a|/a_{\rm ho}$ increases. 
%{\color{red} The derivations result from the perturbation of the non-interacting three-body wavefunction by increasing interaction strength.}
For $^{87}$Rb, since interactions are weak, all the values for $|a|/a_{\rm ho}$ in Fig.~\ref{fig:ER}(a) fall well within the $|a|/a_{\rm ho}\ll1$ regime, the agreement with the LL relation (\ref{gammarM}) holds for all values of $E_q$, including those with energy comparable to $E_{\rm vdW}\approx6000$ kHz$\times h$, therefore, beyond the ultracold regime $E_q\ll E_{\rm vdW}$.
For $^{85}$Rb [Fig.~\ref{fig:ER}(b)], however, due to its strong interactions, even at the lowest trap frequency, the ratio $|a|/a_{\rm ho}$ ($\approx0.56$) is not deep in the $|a|/a_{\rm ho}\ll1$ regime, leading to a deviation from the LL relation for the lowest few $E_q$. For higher excited states, deviations become relatively small as $E_q$ increases, confirming that the LL relation remains valid for $|a|/r_{\rm ho}(q)\ll 1$, as noted previously.   

Also evident in Figs.~\ref{fig:ER}(a) and \ref{fig:ER}(b) is the contrasting behavior of $\Gamma_q$ for $^{87}$Rb and $^{85}$Rb: for $^{87}$Rb, $\Gamma_q$ is strongly dependent on $E_q$, while for $^{85}$Rb, it is relatively independent of $E_q$.
According to the LL relation (\ref{gammarM}), this contrasting behavior can be understood as arising from the distinct energy dependence of $L_3(E)$ for the two isotopes.
For $^{85}$Rb atoms, the primary factor influencing $L_3(E)$ results from its strong interactions ($a=\text{-}5.56$ $r_{\rm vdW}$). In such cases, the expected threshold behavior, $L_3 (E)=\text{const}$, deteriorates for $E\gtrsim E_c=\hbar^2/(ma^2)$, with $L_3(E)$ reaching its unitary behavior, $L_3(E)=C_\eta/E^2$, where $C_\eta=72\sqrt{3}\pi^2\hbar^5/m^3(1-e^{-4\eta})$ and $\eta$ being the three-body inelasticity parameter characterizing, for instance, the Efimov states lifetime and other universal properties of Efimov physics \cite{dincao2004prl,Dincao:2018}. [For $^{85}$Rb, $E_c\approx$ 200 kHz$\times h$ (or 10$\mu$K$/k_B$) and displayed in Fig.~\ref{fig:ER} as a vertical dashed line.]
Figure~\ref{fig:ER}(d) shows this change in behavior in our numerical calculations for $L_3(E)$ (solid line) along with the results for $L_3(E_q)$ obtained from the LL relation (\ref{eq:L3}) (symbols), using our calculated values of $E_q$ and $\Gamma_q$ shown Fig.~\ref{fig:ER}(b).
As can be seen in Fig.~\ref{fig:ER}(d), for $^{85}$Rb, most of the states are in the $E_q>E_c$ regime where $L_3=C_\eta/E_q^2$, in which case the LL relation (\ref{gammarM}) predicts $\Gamma_q=\Gamma_{\rm u} [1-(\hbar\omega_{\rm ho})^2/E_q^2]\simeq \Gamma_{\rm u} [1-1/(2q+3)^2]$, thus consistent with $\Gamma_q$ being weakly dependent on $E_q$ for states beyond $E_q=E_c$ [see Fig.~\ref{fig:ER}(b)].
The unitary width, $\Gamma_{\rm u}=\hbar \omega_{\rm ho}(1-e^{-4\eta})/\pi$, provides a direct mean of extracting $\eta$ from trap systems.
In contrast, for $^{87}$Rb atoms, the interaction is weak ($a=1.21$ $r_{\rm vdW}$) and $L_3(E)$ is expected to be constant within a broader energy range [$E_c$ $\approx$ 4000 kHz$\times h$ (or 200 $\mu$K$/k_B$) for $^{87}$Rb]. However, for $^{87}$Rb, $L_3 (E)$ is enhanced within this region due to the presence of a $d$-wave diatomic shape resonance \cite{wang2012pra}, as shown by our numerical calculations for $L_3$ (solid line) in Fig.~\ref{fig:ER}(c). Therefore, according to the LL relation (\ref{gammarM}), the primary mechanism responsible for the enhancement of $\Gamma_q$ for $^{87}$Rb trapped states in Fig.~\ref{fig:ER}(a) traces back to the presence of the $d$-wave diatomic resonance. We note that the deviations shown in Fig.~\ref{fig:ER}(c) near the position of the $d$-wave resonance are likely due to the couplings from other $\lambda\neq0$ three-body channels, not included in our derivation of the LL relation.

We note that typical experiments with trapped few-atom systems require loading atoms from an ultracold gas into a single trap state. Without a dedicated cooling scheme (see, for instance, Refs.~\cite{zurn2013prl,Kaufman:2015,Goban:2018,Liu:2019,Spence:2022,brooks:2022,Shaw:2025}), atoms will instead be loaded into the trap at some finite temperature, typically $k_BT\gg\hbar\omega_{\rm ho}$. In such cases, if the system is in thermal equilibrium, the LL relation can be extended by averaging Eq.~(\ref{gammarM}) under a thermal distribution of quantum trapped states (see End Matter).

The LL relation (\ref{gammarM}) can be generalized to $N$-atom systems ($N\geq 2$), in which case we obtained
\begin{align}
\Gamma_{N,q}=\frac{1}{C_{N}}&\frac{m^{\frac{3N-3}{2}}}{\hbar^{3N-5}}L_N(E_{N,q})
\nonumber\\
\times&E_{N,q}^{\frac{3N-9}{2}}[E_{N,q}^2-4\alpha_N(\hbar\omega_{\rm ho})^2] \omega_{\rm ho},\label{gammaNm} 
\end{align}
with $C_{N}$=$\pi^{\frac{3N-3}{2}}N^{\frac{5}{2}}2^{\frac{3N-5}{2}}\Gamma[\frac{3N-3}{2}]$ and $\alpha_N$=$(N$-$1)(3N$-$7)(3N$-$5)/64$ (see details and further numerical tests at $N=2$ in Ref.~\cite{sm}). Here, $\Gamma_{N,q}$ and $E_{N,q}$ are the width and energy of the trapped $N$-body state, respectively, and $L_N$ the free-space $N$-body loss coefficient \cite{explainLN,mehta2009prl} for the $J=0$ and $\lambda=0$ partial-wave. For $N>3$, the width of an $N$-body trap state is typically determined by different decay pathways that may involve all or only some of the atoms. In Ref.~\cite{Goban:2018}, for instance, it was found experimentally that for Sr atoms in a $q=0$ state the decay rate for $N=3$, 4, and 5 is approximately proportional to the number of triads, indicating that three-body processes remain the dominant decay pathway. In reality, the direct experimental observable is the total decay rate, including all higher-order processes, defined as,
\begin{align}
    \frac{\Gamma_{N,q}^{\rm tot}}{\hbar}&=\sum_{K=3}^{N} \binom{N}{K}\frac{\Gamma_{K,q}}{\hbar}\nonumber\\
    &=\binom{N}{3}\frac{\Gamma_{3,q}}{\hbar}+\binom{N}{4}\frac{\Gamma_{4,q}}{\hbar}+\cdots+\frac{\Gamma_{N,q}}{\hbar},\label{GammaTot}
\end{align}
with fundamental few-body decay rates $\Gamma_{K,q}/\hbar$ given in Eq.~(\ref{gammaNm}).
In an experiment such as Ref.~\cite{Goban:2018}, 
where trap states containing $N=3,4,5,\ldots$ atoms are prepared 
and their decay rates ${\Gamma_{N,q}^{\rm tot}}/{\hbar}$ measured, 
the corresponding fundamental rates can be extracted 
by inverting Eq.~(\ref{GammaTot}), 
\begin{align}
    \frac{\Gamma_{N,q}}{\hbar}&=\sum_{K=3}^{N}(-1)^{N-K}\binom{N}{K}\frac{\Gamma_{K,q}^{\rm tot}}{\hbar},
    \label{GammaPatial}
\end{align}
allowing for the determination of $N$-body scattering rates $L_N$ via the LL relation (\ref{gammaNm}), in a much more controlled way than in ultracold gas experiments \cite{explainLN,kraemer2006NT,stecher2009NP,ferlaino2009PRL,zenesini2013NJP}. 

In summary, we have derived the Lellouch-L\"uscher (LL) relation connecting the energy and width of few-body states in a harmonic trap to the corresponding loss constants in free space, and demonstrated its validity through three-body numerical calculations. A unique feature of the LL relation is that, when atoms are prepared in trap states of a single partial wave---typically $J=0$ for bosons---it enables the measurement of individual partial-wave contributions to loss rates. This makes trapped few-body systems a more sensitive probe of scattering resonances and interference phenomena, including those associated with Efimov physics \cite{Dincao:2005,Dincao:2018}, which are often obscured in bulk-gas measurements due to thermal averaging over multiple partial waves \cite{dincao2004prl}. We further show that the generalized $N$-body LL relation provides a systematic path to precise determinations of multi-body scattering rates \cite{Goban:2018,mehta2009prl}. 
Taken together, the results from the LL relation framework help to establish trapped few-body systems as a powerful platform for advancing our understanding of complex quantum phenomena.

\textit{Acknowledgments}—This work was supported by the Baden-W\"urttemberg Stiftung through the Internationale Spitzenforschung program 
(BWST, contract No.~ISF2017-061) and by the German Research Foundation (DFG, Deutsche Forschungsgemeinschaft, 
contract No. 399903135). The authors acknowledge support by the state of Baden-Württemberg through bwHPC and the German Research Foundation (DFG) through grant no INST 40/575-1 FUGG (JUSTUS 2 cluster). J.H.D and J.P.D. acknowledge funding by Q-DYNAMO (EU HORIZON-MSCA-2022- SE-01) within project No. 101131418. J.P.D. also acknowledges partial support from the U.S. National Science Foundation (PHY-2308791/PHYS-2452751) and the Office of Naval Research (ONR) grant N00014-21-1-2594.

\textit{Data availability}—The data that support the findings of this article are openly available \cite{data}
\clearpage
\newpage
\onecolumngrid
 \vspace{1em}
\begin{center}
    \large\bfseries End Matter
\end{center}
 \vspace{0.5em}
\twocolumngrid

\appendix
 \section{LL relation in thermal equilibrium}   
 In experiments, loading a few atoms from the bulk into an isolated trap generally leads to $k_BT\gg\hbar\omega_{\rm ho}$, producing a thermal ensemble instead of a single trap state. The corresponding lifetime $\hbar/\Gamma_{\rm th}$ of the thermal state can be estimated by assuming a Boltzmann distribution of energies $E_q$ as
 \begin{eqnarray}
 \Gamma_{\rm th} (T)\equiv\frac{\sum_{q}e^{-E_q/k_B T} \Gamma_q}{\sum_{q}e^{-E_q/k_B T}}. \label{gammath}
 \end{eqnarray}
 Provided that the initial state (prior to loading) has a low enough temperature ($k_BT_{\rm in}\ll E_{c}$) the populated trap states will predominantly be that of the $J=0$ (and $\lambda=0$) partialwave. This allows us to use the LL relations [Eq.~(\ref{gammarM}) or (\ref{gammaNm})] to derive simple results in two limiting cases, namely, the threshold ($|k_{\rm th}a|\ll1$) and unitary ($|k_{\rm th}a|\gg1$) regimes, with $\hbar k_{\rm th}=(m k_BT)^{1/2}$. In fact, while in the threshold regime most of the states that contribute to Eq.~(\ref{gammath}) are those with $\hbar\omega_{\rm ho}\ll E_q\sim k_BT \ll E_c$, where $L_3(E_q)\approx L_3(0)$, the states contributing to Eq.~(\ref{gammath}) in the unitary regime will be those with $\hbar\omega_{\rm ho}\ll E_c\ll E_q\sim k_BT$, where $L_3(E_q)=C_{\eta}/E_q^2$. Assuming $E_q\simeq(2q+3)\hbar\omega_{\rm ho}$ and using the LL relation (\ref{gammarM}), the thermal loss rate in Eq.~(\ref{gammath}) can be expressed as
 \begin{align}
  \Gamma_{\rm th}=
     \begin{cases}
         2\frac{1}{C}\frac{m^3}{\hbar^4} L_3(0)(k_B T)^2\omega_{\rm ho}, & \mbox{for $|k_{\rm th}a|\ll1$,}\\
        (1-e^{-4\eta})\hbar\omega_{\rm ho}/\pi=\Gamma_{\rm u}, & \mbox{for $|k_{\rm th}a|\gg1$,}
     \end{cases}\label{gammathX}
 \end{align}
for which a brief derivation and numerical test will be presented below.
We note that for $|k_{\rm th}a|\ll1$ the thermal decay rate is quadratic in $(k_BT)$ and is a factor 2 larger than the rate obtained by the LL relation (\ref{gammarM}), for $E_q=k_BT$ and $L_3(E_q)\simeq L_3(0)$. For $|k_{\rm th}a|\gg1$, however, the thermal decay rate is independent of the temperature. This unexpected result is a direct consequence of the LL relation (\ref{gammarM}). Similar results can be obtained for $N$ atoms in thermal equilibrium using the total decay rate in Eq.~(\ref{GammaTot}) and the corresponding LL relations (\ref{gammaNm}).

\begin{figure}[t]
 \centering
  \resizebox{0.49\textwidth}{!}{\includegraphics{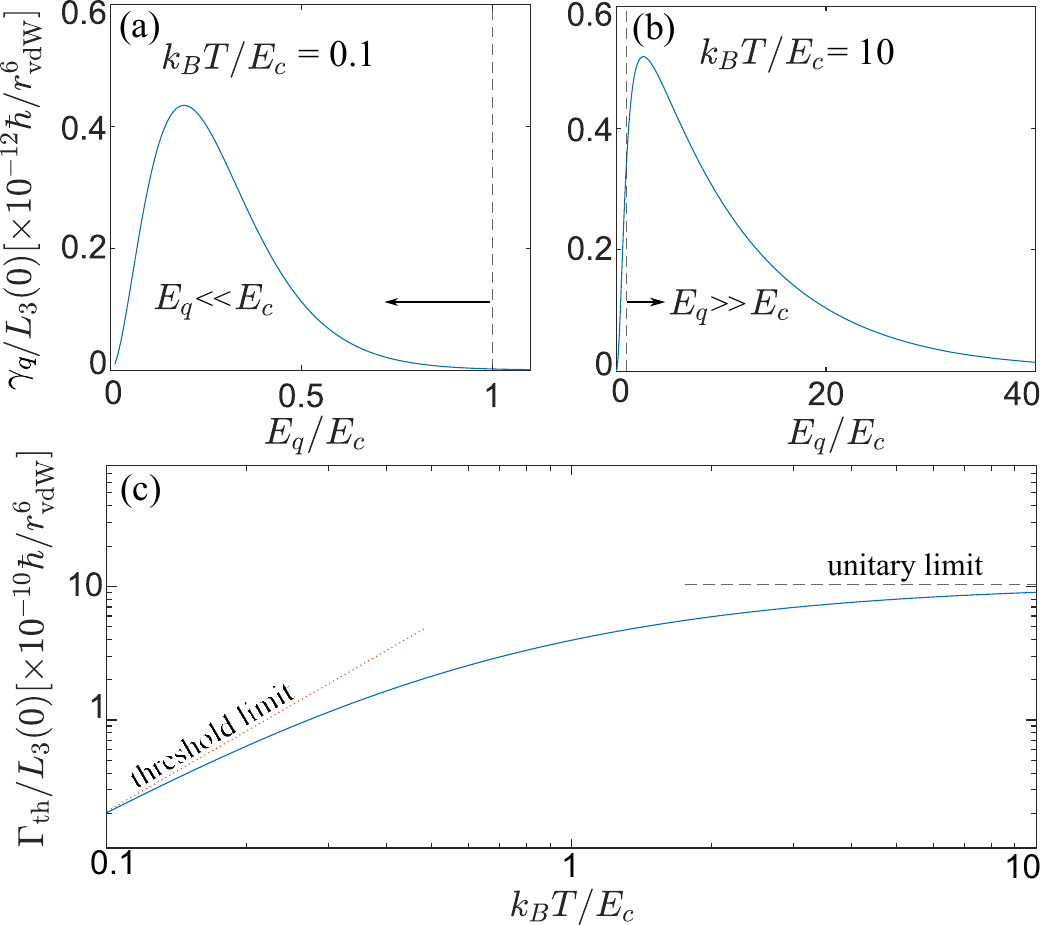} }
 \caption{\label{fig:tem} The distribution of the individual contribution, $\gamma_q$, of the trap state to $\Gamma_{\rm th}$ in the threshold regime of $k_BT/E_c=0.1$ (a) and in the unitary regime of $k_BT/E_c=10$ (b). The vertical dashed lines indicate that $E_q=E_c$. (c) shows the temperature-dependence behavior of the $\Gamma_{\rm th}$ calculated from Eq. (\ref{gammath}) and from Eq.~(\ref{gammathX}) at threshold limit (dotted line) and unitary limit (dashed line). The displayed result is obtained by using Eq. (\ref{eq:L3toy}) for $L_3(E)$ with $\alpha=1$, $E_c=0.1 E_{\rm vdW}$ and $\hbar \omega_{\rm ho}=0.001 E_{\rm vdW}$ and Eq. (\ref{gammarM}) for $\Gamma_q$, for instance.}
\end{figure}
To derive Eq. (\ref{gammathX}), we define the individual contribution of the $q$th trapped state to $\Gamma_{\rm th}$ as 
\begin{equation}
\gamma_q= 
\frac{e^{-E_q/k_BT}\Gamma_q}{\sum_q e^{-E_q/k_BT}},
\end{equation}
such that $\Gamma_{\rm th}=\sum_q \gamma_q$. Due to the Boltzmann factor $e^{-E_q/k_BT}$, temperature acts as a natural cutoff: a state will contribute significantly to $\Gamma_{\rm th}$ only if its energy $E_q$ is of the order of $\sim k_BT$ or smaller. Consequently, in the threshold regime where $k_BT\ll E_c=\hbar^2/ma^2$, the major contributing states have energies $\hbar\omega_{\rm ho}\ll E_q \ll E_c$. This is demonstrated in Fig. \ref{fig:tem}(a) by using the following toy model for $L_3(E)$
\begin{align}\label{eq:L3toy}
    L_3(E)=\frac{L_{3}(0)}{1+\alpha(E/E_c)^2},
\end{align}
where $\alpha=L_3(0)E_c^2/C_{\eta}$. [This toy model is designed to capture the two limiting behaviors of 
$L_3(E)$: $L_3(E)\rightarrow L_3(0)$ as $E \rightarrow 0$ (threshold limit) and $L_3(E) \rightarrow C_{\eta}/E^2$ as $ E \to \infty$ (unitary limit),
thus providing a suitable framework for the analysis in this section.]

As the majority contribution to $\Gamma_{\rm th}$ comes from the states with $\hbar \omega_{\rm ho} \ll E_q\ll E_c$, we are allowed to replace the energy dependence $L_3(E)$ with $L_3(0)$ and $[E_q^2-\hbar^2\omega_{\rm ho}^2]$ with $E_q^2$ in Eq. (\ref{gammath}). By further assuming $E_q \simeq (2q+3)\hbar \omega_{\rm ho}$, we get 
\begin{equation}
\Gamma_{\rm th}=\frac{1}{C}\frac{m^3}{\hbar^2} L_3(0)  \omega_{\rm ho}^3 g\left(\frac{\hbar\omega_{\rm ho}}{k_B T}\right) ,
\end{equation}
with
\begin{eqnarray}
g(x)&=&5+4\coth(x)+2\text{csch}^2(x) \\ \notag
    &=&\frac{2}{x^2}+\frac{4}{x}+\frac{13}{3}+O(x),
\end{eqnarray}
where the leading order gives Eq. (\ref{gammathX}) at $|k_{\rm th}a|\ll1$. The validity of our Eq. (\ref{gammathX}) is demonstrated in Fig. \ref{fig:tem}(c), where it is compared to the numerical result using the expression (\ref{eq:L3toy}) for $L_3(E)$.

At the unitary limit where $k_BT\gg E_c$, the states that contribute to $\Gamma_{\rm th}$ still have an $E_q$ of the order of $\sim k_BT$ or smaller. However, in this regime, a majority of these states can satisfy $E_q \gg E_c \gg \hbar \omega_{\rm ho}$ (see Fig. \ref{fig:tem}(b)), such that $L_3(E_q) = C_{\eta}/E_q^2$. To further simplify the derivation, we assume $C_{\eta}=L_3(E_q)E_q^2$ for all states. Plugging this expression into Eq. (\ref{gammath}) and by considering $[E_q^2-\hbar^2\omega_{\rm ho}^2]\rightarrow E_q^2$, we get the simple relation in Eq.~(\ref{gammathX}), i.e., $\Gamma_{\rm th}=\Gamma_{\rm u}$. The validity of Eq. (\ref{gammathX}) in this limit is also confirmed in Fig. \ref{fig:tem}(c) by comparing it to the numerical result from the toy model.

\clearpage
\newpage
\section{Supplemental Material}
\subsection{Hyperspherical Adiabatic Representation}

%\add{JPD: needs to describe $W_\lambda$ and $W_f$ potentials.}

In the hyperspherical adiabatic representation, the total wave function $\Psi(R,\Omega)$ of three atoms can be expressed as \cite{Suno:2002,wang2011pra}:
\begin{align}
    \Psi(R,\Omega)=\sum_{\nu}F_{\nu}(R)\Phi_\nu(R;\Omega),
\end{align}
where $F_{\nu}(R)$ is determined by the hyperradial equation (2) in the main text and $\Phi_{\nu}(R;\Omega)$ are the eigenfunctions of the adiabatic hyperangular equation:
\begin{equation} \label{Schroa}
\hat{H}_{\rm ad}\Phi_{\nu}(R;\Omega)=U_{\nu}(R)\Phi_{\nu}(R;\Omega)
\end{equation}
at fixed $R$. Here $\Omega$ denotes a set of 5 hyperangles (including Euler angles) describing the internal and overall rotational motion of the three-atom system.

To solve the adiabatic hyperangular equation (\ref{Schroa}), we use the same interaction model as in Ref. \cite{Lipaper}, which we briefly introduce below. The adiabatic Hamiltonian $\hat{H}_{\rm ad}$,
\begin{eqnarray}
\hat{H}_{\rm ad}=\frac{\hat\Lambda^2(\Omega)+15/4}{2\mu R^2}\hbar^2+\hat{V}_T(R,\Omega)+\hat{H}_{\rm hf+Z}(B),
\end{eqnarray}
contains the hyperangular kinetic energy via the hyperangular momentum operator \cite{Suno:2002,wang2011pra}, 
$\hat\Lambda$, the hyperfine and Zeeman Hamiltonian, $\hat{H}_{\rm hf+Z}$, as well as all the interatomic interactions, $\hat{V}_{T}(R,\Omega)=\hat{V}(r_{12})+\hat{V}(r_{23})+\hat{V}(r_{31})$. We note that the Zeeman interaction disappears in the present study as we consider $B=0$. 
The pairwise interatomic interaction, $\hat{V}(r_{ij})=\sum_{SM_S}|SM_S\rangle V_S(r_{ij})\langle SM_S|$, is represented by the realistic singlet ($S=0$) and triplet ($S=1$) Bohr-Oppenheimer 
potentials, where $S$ is the total electronic spin and $M_S$ is its projection quantum number. At long range, both singlet and triplet potentials are dominated by the van der Waals interaction, characterized by the length $r_{\rm vdW}=(mC_6/\hbar^2)^{1/4}/2$ and energy $E_{\rm vdW}=\hbar^2/mr_{\rm vdW}^2$ scales, where $C_6$ is the van der Waals dispersion coefficient \cite{Chin:2010}. In this work, we focus on $^{85}$Rb$_2$ and $^{87}$Rb$_2$, for which the characteristic length scale are $r_{\rm vdW}\approx82.16 a_0$ and $r_{\rm vdW}\approx82.64 a_0$, respectively~\cite{Lipapertwo}. Here, $a_0$ denotes the Bohr radius. To reduce the computational burden, we add a short-range repulsive term for each potential to control the number of bound states in the system \cite{Lipapertwo}. As a result, each interaction potential has 6 $s$-wave bound states except that the triplet potential of $^{85}$Rb$_2$ has 5 $s$-wave bound states. Nevertheless, the reduced singlet and triplet potentials successfully reproduce the weakly bound states and the low-energy scattering properties. 

The solutions of the hyperangular adiabatic equation are obtained by expanding the channel functions $\Phi_{\nu}$ on the basis of 
separated atoms hyperfine spins states, $|\sigma\rangle\equiv|f_{i}m_{f_i}\rangle|f_{j}m_{f_j}\rangle|f_{k},m_{f_k}\rangle$,
\begin{align}
    \Phi_\nu(R;\Omega)=\sum_{\sigma}\phi^{\sigma}_{\nu}(R;\Omega)|\sigma\rangle,
\end{align}
where $f$ and $m_f$ denote the atomic hyperfine and its magnetic projection quantum numbers, respectively. In practice, we fix the third ($k$) atom's spin at its initial state, i.e., $|f_{k},m_{f_k}\rangle=|f_{k}^{\rm in},m_{f_k}^{\rm in}\rangle$.
We note that the validity of fixing the third atom's spin has been demonstrated for Rb in Ref. \cite{Lipaper}. In this work, the initial atomic spin state is chosen as $|f_{k}^{\rm in},m_{f_k}^{\rm in}\rangle=|2,\text{-}2\rangle$ for $^{85}$Rb and $|1,\text{-}1\rangle$ for $^{87}$Rb, respectively.
Applying this expansion to the hyperangular adiabatic equation results in a coupled system of equations for the 
components of $\phi_\nu^{\sigma}$:
\begin{eqnarray}
\left[\frac{\hat\Lambda^2(\Omega)+15/4}{2\mu R^2}\hbar^2+E^{\sigma}_{\rm hf}(B)-U_{\nu}(R)\right]\phi^{\sigma}_{\nu}(R;\Omega)\nonumber \\
+\sum_{\sigma'}V_T^{\sigma\sigma'}(R,\Omega)\phi^{\sigma'}_{\nu}(R;\Omega)=0,\label{FullAngEq}
\end{eqnarray}
where $E_{\rm hf}^{\sigma}$ is the sum of the hyperfine energies of the three separated atoms in the $|\sigma\rangle$ spin state at the magnetic field $B$. 
After Eq.~(\ref{FullAngEq}), the adiabatic potentials $W_{\nu}$ and non-adiabatic couplings $W_{\nu \nu'}$ can be obtained for the use in hyperradial equation~(2)~\cite{Suno:2002,wang2011pra}. The adiabatic potentials have two possible asymptotic ($R\rightarrow \infty$) forms:
    \begin{align}
        W_{\nu}(R)=W_{f}(R)&\simeq -E_b+\frac{l(l+1)}{2\mu R^2}\hbar^2,\\
        W_{\nu}(R)=W_{\lambda}^b(R)&\simeq E_{\rm hf}^{\sigma}+\frac{\lambda(\lambda+4)+15/4}{2\mu R^2}\hbar^2, \label{eq:wc}
    \end{align}
defining the atom-molecule decay channels (labelled by $f$) and three-atom continuum channels (labelled by $\lambda$), respectively. Here, $E_b$ denotes the molecular binding energy. In the present work, the total number of atom-molecule decay channels is 112 for both $^{85}$Rb and $^{87}$Rb, including molecular states with all possible spins and partial waves. The continuum three-body channels are included up to several tens to ensure good convergence.

For three atoms in an isotropic harmonic trap ($\omega_x = \omega_y = \omega_z=\omega_{\rm ho}$), adding the trap in our model only affects the hyperradial motion, leaving the hyperangular equation~(\ref{Schroa}) unchanged. Therefore, the effect of the isotropic harmonic trap is captured by adding $\frac{1}{2}\mu \omega_{\rm ho}^{2}R^2$ to the free-space equation (2). For numerical implementation, it is both practical and physically reasonable to include the trap potential only in the three-atom continuum channels $W_{\lambda}$, while leaving all atom–dimer channels $W_f$ unconfined to allow decays \cite{Goban:2018,mark2020PRR}. This choice is motivated by the fact that the trap depth is typically much smaller than the kinetic energy released during molecule formation via decay pathways. As a result, the influence of the trap on the recombination dynamics within $W_f$ channels is negligible.
The trapped state is then obtained from a standard atom–dimer scattering calculation in the relevant energy regime, where the resonance feature is identified. 
\subsection{Reactive three-body probability density $\mathcal{D}_q$}
 \begin{figure}[t]
 \centering
  \resizebox{0.5\textwidth}{!}{\includegraphics{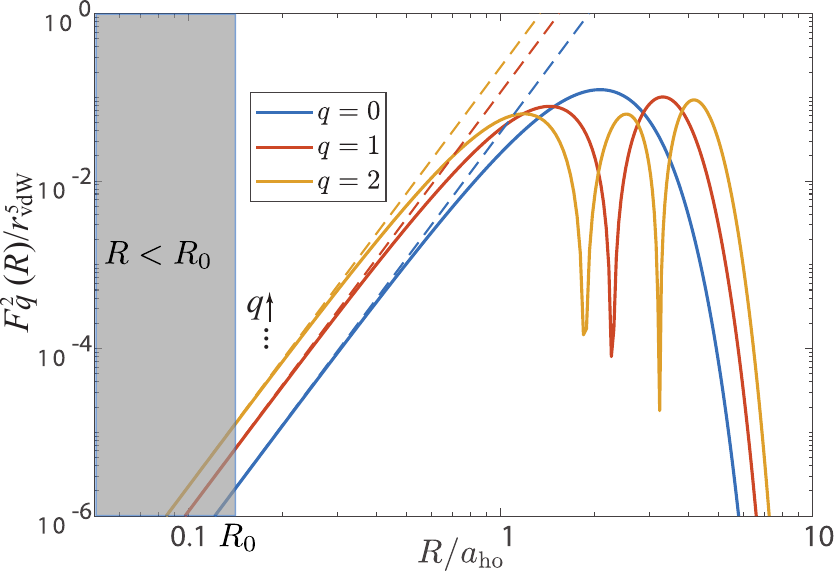} }
 \caption{\label{fig:psr} Non-interacting three-body hyperradial wavefunction $F_q(R)$ in a spherical harmonic trap. The solid lines show $F_q^2(R)$ from Eq. (7), while the dashed lines represent the corresponding Taylor expansion to leading order. The shaded area indicates the reaction regime $R<R_0$.
 }
\end{figure}
In our studies, we define the reactive three-body probability density $\mathcal{D}_q$ as:
\begin{eqnarray} \label{pqdef}
\mathcal{D}_q&\equiv&\frac{1}{\mathcal{V}_0}\int_{\mathcal{V}_0}|\Psi_{3,q}(\vec{R}_{\rm cm},\vec{R})|^2d^3\!\vec{R}_{\rm cm}d^6\!\vec{R}, \\ \notag
%&=&\frac{1}{\pi^3R_0^6}\int_{|\vec{R}|<R_0}|\Psi_{q}(\vec{R})|^2d\vec{R}_{\rm %cm}d\vec{R} \\ \notag
\end{eqnarray}
by averaging the probability of the wavefunction $\Psi_{3,q}(\vec{R}_{\rm cm},\vec{R})$ for $q$-th trapped three-body state over the hyperspherical volume $\mathcal{V}_0=\int d^3R^5dR=\pi^3R_0^6/6$ ($d^6\!\vec R=R^5dRd^5\Omega$) within the reaction regime ($|\vec{R}|<R_0$). The parameter $R_0$, characterizing the recombination regime, is expected to be of the order of the interaction range. For van der Waals interaction, for instance, we consider $R_0\sim r_{\rm vdW}=\frac{1}{2}(mC_6\hbar^2)^{1/4}$, where $C_6$ is the dispersion coefficient. Here, $\Psi_{3,q}(\vec{R}_{\rm cm},\vec{R})$ describes both the 3D center of mass motion in $\vec{R}_{\rm cm}$ and 6D relative motion in $\vec{R}=(R,\Omega)$. We note that the center of mass motion is not essential for defining $\mathcal{D}_q$ as it does not play a role in the reaction process. Nevertheless, we keep it in our definition (\ref{pqdef}) for completeness. In the hyperspherical coordinate representation, we consider only the incoming $\lambda=0$ channel to express the trapped three-body wavefunction as $\Psi_{3,q}(\vec{R}_{\rm cm},\vec{R})=\Psi_{\rm cm}(\vec{R}_{\rm cm})\Phi_0(R;\Omega)F_q(R)R^{-5/2}$, normalized by $\int |\Psi_{\rm cm}(\vec{R}_{\rm cm})|^2d^3\vec{R}_{\rm cm} =1$, $\int |\Phi_0(R;\Omega)|^2 d\Omega=1$ and $\int |F_q(R)|^2 dR =1$. By using $d^3 \vec{R}=R^5dRd\Omega$ and integrating over $\vec{R}_{\rm cm}$ and $\Omega$ in Eq. (\ref{pqdef}), we get 
\begin{equation} \label{pqr}
\mathcal{D}_q=\frac{1}{\mathcal{V}_0}\int_0^{R_0}F_{q}(R)^2dR.
\end{equation}
We then apply the Taylor expansion around ($R=0$) to the non-interacting three-body trap-state wavefunction $F_q^2(R)$ [given by Eq.~(7) of the main text]:
\begin{equation} \label{fqtaylor}
F_{q}^2(R)=\frac{R^5}{N_qa_{\rm ho}^6}\left[\left(\frac{\Gamma(q+3)}{q!\Gamma(3)}\right)^2+O\left(\frac{R^2}{a_{\rm ho}^2}\right)\right].
\end{equation}
We neglect the $O({R^2}/a_{\rm ho}^2)$ terms, which is justified by the assumption $R_0 \sim r_{\rm vdW} \ll a_{\rm ho}$ (see, for example, Fig. \ref{fig:psr}). As a result, we get to leading order
\begin{eqnarray} \label{psr2}
\mathcal{D}_q&=&\frac{12\int_0^{R_{\rm 0}}R^5dR}{\pi^3R_0^6(q+1)(q+2)3\sqrt{3}a_{\rm ho}^6}\left(\frac{\Gamma(q+3)}{q!\Gamma(3)}\right)^2 \\ \notag 
&=&\frac{(q+1)(q+2)}{6\sqrt{3}\pi^3a_{\rm ho}^6}  
=\frac{m^3\omega_{\rm ho}(E_q^2-\hbar^2\omega_{\rm ho}^2)}{24\sqrt{3}\pi^3\hbar^5},
\end{eqnarray} 
where we make the substitution $(2q+3)\hbar\omega_{\rm ho}$$\rightarrow$$E_{q}$, corresponding to the energy of three non-interacting atoms, $(2q+3)\hbar\omega_{\rm ho}$.
In the weakly interacting regime $|a|/a_{\rm ho} \ll 1$, where the interaction-induced shifts in the energy, $E_q$, and size, $r_q$, of the states can be treated perturbatively as a power series in $a/a_{\rm ho}$,
\begin{align}
    &E_q/\hbar\omega_{\rm ho} = (2q+3) + {\cal O}({a}/{a_{\rm ho}}),\\
    &r_q/a_{\rm ho} = (2q+3)^{1/2} + {\cal O}({a}/{a_{\rm ho}}),
\end{align}
the substitution $(2q+3)\hbar\omega_{\rm ho} \rightarrow E_q$ remains valid, since $(2q+3)\hbar\omega_{\rm ho}$ is 
still an excellent approximation to the actual trapped-state energy $E_q$. As interactions increase toward the 
regime $|a|/a_{\rm ho} \sim 1$, however, the trapped eigenstates acquire measurable shifts in both energy and 
spatial extent. Since $\mathcal{D}_q$ quantifies the probability density within the short-range reaction region, it 
becomes sensitive to these interaction-induced modifications of the size of the trapped state. As shown in previous studies in the two-body sector 
\cite{Bolda:2002,Kohler:2005}, in the 
non-perturbative regime $|a|/a_{\rm ho} \lesssim 1$, the replacement of $(2q+3)\hbar\omega_{\rm ho}$ by the actual 
energy $E_q$ captures the leading order effect of the modification of the size of the state 
due to interaction, leading to 
the renormalization of $\mathcal{D}_q(E_q)$ as given in Eq. (\ref{psr2}).
%\textcolor{red}{Similar to previous studies in the two-body sector \cite{Bolda:2002,Kohler:2005},
%replacing $(2q+3/2)\hbar\omega_{\rm ho}$ by the interacting 
%energies $E_{q}$ accounts for interaction-induced modifications of the 
%effective reaction volume.}

%In this sense, the substitution $(2q+3)\hbar\omega_{\rm ho}$$\rightarrow$$E_{q}$ accounts 
%for interaction-induced modifications of the effective reaction volume.}

\begin{figure}[t]
 \centering
  \resizebox{0.48\textwidth}{!}{\includegraphics{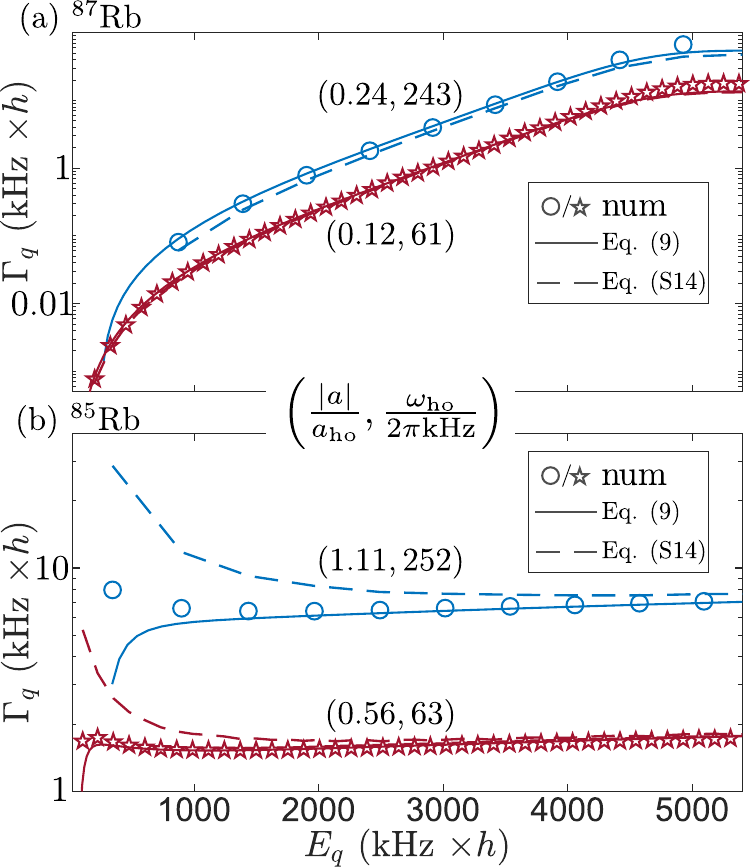} }
 \caption{\label{fig:add} Comparison of $\Gamma_q$ obtained from the numerical calculation (symbols) and from the LL relation of Eq. (9) in the main text (solid lines) to that obtained from LL relation of Eq. (\ref{gammarMni}) without the substitution $(2q+3)\hbar\omega_{\rm ho}\rightarrow E_{q}$ in $\mathcal{D}_q$ (dashed lines) for $^{87}$Rb (a) and $^{85}$Rb (b). The numbers in the parentheses indicate the trap frequency and the ratio $|a|/a_{\rm ho}$. For both results from the numerical calculation and from Eq. (\ref{gammarMni}), the data points correspond to quantized trap states, starting from $q=0$.
 }
\end{figure}

In fact, in our studies we found that using the actual interacting energies $E_q$ the resulting LL-relations 
[Eqs.~(1) and (9) of the main text] indeed lead to a better agreement with our fully numerical calculations in 
the $|a|/a_{\rm ho}\sim1$ regime. Below, we compare our results with that obtained without using the substitution $(2q+3)\hbar\omega_{\rm ho}\rightarrow E_{q}$ substitution in $\mathcal{D}_q$ [Eq. (\ref{psr2})], in which case the resulting LL-relation is expressed as,
\begin{equation}
\Gamma^{\rm ni}_q=\frac{1}{C}\frac{m^3}{\hbar^4} L_3(E_q)[(2q+3)^2-1]\hbar^2\omega_{\rm ho}^3. \label{gammarMni}
\end{equation}
Note that in this case, for each value of $q$ the effect of interactions is only incorporated through the values of $L_3$ calculated at 
$E=E_q$. In Figs. \ref{fig:add}(a) and \ref{fig:add}(b) we show our numerically calculated values of $\Gamma_q$ for 
$^{87}$Rb and $^{85}$Rb, respectively, with those obtained from $\Gamma_q^{\rm ni}$ [Eq.~(\ref{gammarMni})] and 
the from $\Gamma_q$ [Eq. (9) of main text]. 
[Although Eq.~(\ref{gammarMni}) is evaluated only for discrete values of $q$, the results in Fig.~\ref{fig:add} are represented as lines connecting the calculated values.] From Fig.~\ref{fig:add}, it is clear that as the interaction increases from perturbative 
($a/a_{\rm ho}\ll 1$) [see Fig.~\ref{fig:add}(a) for $^{87}$Rb] to non-perturbative ($|a/a_{\rm ho}|\sim1$) 
[Fig.~\ref{fig:add}(b) for $^{85}$Rb], the agreement between the numerical values of $\Gamma_q$ and $\Gamma^{\rm 
ni}_q$ quickly deteriorates. The agreement with the numerical calculation is substantially improved when the LL-relation (9) for $\Gamma_q$ is employed. We note that the 
agreement between all models for $\Gamma_q$ becomes increasingly better for higher excited trap states with the 
increase of $q$. This, however, relates to the fact that as $q$ increases so does the size of excited states, now 
denoted by $r_{\rm ho}(q)\simeq(2q+3)^{1/2}a_{\rm ho}$, thus effectively reducing the effect of the interactions in 
the reaction volume implicit in $\mathcal{D}_q$. In this sense, $|a|/r_{\rm ho}(q)$ provides a more accurate 
characterization of the interaction effects for individual trap states.

We note that a more rigorous treatment of the non-perturbative regime would require 
replacing the non-interacting hyperradial wave function [Eq.~(7) of the main text] with 
the proper interacting wave function in order to determine the corresponding changes in $\mathcal{D}_q$ [Eq.~(\ref{psr2})] and the LL relations. Such an extension becomes essential in the strongly interacting regime, $|a|/a_{\rm ho}\gg1$. However, incorporating these effects would significantly complicate the otherwise simple formalism developed here, which is intended for the regime $|a/a_{\rm ho}|\lesssim1$. Future work will focus on incorporating interaction effects through the approach outlined above.

 According to Eq. (\ref{psr2}), the reactive three-body probability density increases with the trap frequency $\omega_{\rm ho}$ as a higher $\omega_{\rm ho}$ typically compresses the occupied volume and raises the particle density. For a given $a_{\rm ho}$, however, the reactive three-body probability density $\mathcal{D}_q$ increases with the energy of the trap state $E_q$ (or equivalently with the state index $q$), even though the state size expands as $r_{\rm ho}(q)\simeq(2q+3)^{1/2}a_{\rm ho}$. This behavior arises because the wavefunction amplitude at short range ($R\ll a_{\rm ho}$) increases with the trap state index $q$, as is shown in Fig. \ref{fig:psr}. At the same time, Fig. \ref{fig:psr} also shows that the wavefunction amplitude grows at long range ($R\gg a_{\rm ho}$) with an increasing $q$, leading to an overall increase in the state size. That is to say, as
$q$ increases, the wavefunction amplitude in the trap state is pushed toward both short- and long-range regions, resulting in a higher reactive probability density $\mathcal{V}_0$ as the trap state expands.
%This also explains the decrease in effective volume $V_{\rm eff}$ (see main text) %with an increasing trap state size.

Note that the short-range parameter $R_0$ does not appear in Eq. (\ref{psr2}) because it cancels between the integration contribution and $\mathcal{V}_0$ in Eq. (\ref{pqr}). Consequently, the precise choice of $R_0$ will not affect the physical result. Practically, $R_0=\beta r_{\rm vdW}$ can be chosen to scale with the characteristic length scale, $r_{\rm vdW}$, of the van der Waals interaction between atoms. A reasonable value for $\beta$ is considered to be $1\leq \beta \ll \frac{a_{\rm ho}}{r_{\rm vdW}}$.

\subsection{Connection $\mathcal{D}_q$ to the three-body correlation function}
%\add{In order to make connection to the near-threshold relation $\Gamma/\hbar = L_3\int d^3\vec{r}|\phi_{0}(r)|^{6}/3$, we rewrite the Eq. (\ref{pqdef}) as:
In order to explore the relationship between the reactive three-body probability density 
$\mathcal{D}_q$ and the three-body correlation function, we rewrite Eq. (\ref{pqdef}) as:
%For our purpose of this subsection, we rewrite Eq. (\ref{pqdef}) as:
\begin{equation} \label{pqre}
\mathcal{D}_q=\frac{1}{\pi^3}\int I_{R_0}(\vec{R}_{\rm cm},\Omega)d^3\!\vec{R}_{\rm cm}d^5\Omega
\end{equation}
with 
\begin{equation}
I_{R_0}(\vec{R}_{\rm cm},\Omega)=\frac{6}{R_0^6}\int_0^{R_0}|\Psi_{3,q}(\vec{R}_{\rm cm},\vec{R})|^2R^5dR.
\end{equation}
Building on the fact that the three-body correlation function is defined in the context of contact interactions \cite{Reynolds:2020}, we take the limit $R_0\rightarrow 0$, yielding
\begin{eqnarray}
I_{R_0\rightarrow 0}(\vec{R}_{\rm cm},\Omega)&=&|\Psi_{3,q}(\vec{R}_{\rm cm},\vec{R})|^2_{R=0}  \notag \\
&=&\int|\Psi_{3,q}(\vec{R}_{\rm cm},\vec{R})|^2 \delta(R)dR.
\end{eqnarray}
By plugging the above expression of $I_{R_0\rightarrow 0}(\vec{R}_{\rm cm},\Omega)$ into Eq. (\ref{pqre}), we get 
\begin{align} \label{pqlim}
\lim_{R_0 \to 0} \mathcal{D}_q&=\frac{1}{\pi^3}\int |\Psi_{3,q}(\vec{R}_{\rm cm},\vec{R})|^2\delta(R)d^3\!\vec{R}_{\rm cm}dRd\Omega \notag \\
&=\int |\Psi_{3,q}(\vec{R}_{\rm cm},\vec{R})|^2 \delta(\vec{R})d^3\!\vec{R}_{\rm cm}d^6\!\vec{R}  \notag \\
&=\int |\Psi_{3,q}(\vec{R}_{\rm cm},\vec{R})|^2 \delta^{(3)}(\vec{\rho}_{23})
\delta^{(3)}(\vec{\rho}_1) \notag \\
&\quad\quad\quad\quad\quad\quad\quad\quad\times d^3\!\vec{R}_{\rm cm}d^3\!\vec{\rho}_{23}d^3\!\vec{\rho}_1, 
\end{align}
where we use $\delta(\vec{R})=\delta(R)/(\pi^3R^5)$ and $d^6\!\vec{R}=R^5dRd\Omega$ to get the second line. The 6D position vector $\vec{R}=(R,\Omega)$ in relative motion can also be decomposed in Jacobi coordinates $\vec{R}=(s_k^{-1}\vec{\rho}_{ij},s_k\vec{\rho}_{k})$ with $i\neq j \neq k$, where $\vec{\rho}_{ij}=\vec{r}_i-\vec{r}_j$, $\vec{\rho}_{k}=\vec{r}_k-(\vec{r}_i+\vec{r}_j)/2$, $s_i=s_j=s_k=s=2^{1/2}/3^{1/4}$ for identical particles \cite{Dincao:2018}. From the second to the third line of the above equation, we take $\vec{R}=(s^{-1}\vec{\rho}_{23},s\vec{\rho}_{1})$, for instance, which results in $\delta(\vec{R})=\delta(s^{-1}\vec{\rho}_{23})\delta(s\vec{\rho}_{1})=\delta(\vec{\rho}_{23})\delta(\vec{\rho}_{1})$ and $d\vec{R}=d(s^{-1}\vec{\rho}_{23})d(s\vec{\rho}_{1})=d\vec{\rho}_{23}d\vec{\rho}_{1}$. Here $\vec{r}_1, \vec{r}_2, \vec{r}_3$ denote the single-atom 3D position vectors, where $\vec{R}_{\rm cm}$ is given by $\vec{R}_{\rm cm}=(\vec{r}_1+\vec{r}_2+\vec{r}_3)/3$. Next, we take $q=0$ and make the following transformation on the three-body total wavefunction 
\begin{eqnarray} \label{psitran}
&&\Psi_{3,q=0}[\vec{R}_{\rm cm},\vec{R} \rightarrow (s^{-1}\vec{\rho}_{23},s\vec{\rho}_{1})] =\langle\vec{R}_{\rm cm},\vec{\rho}_{23},\vec{\rho}_{1}|\Psi_{3,0}\rangle \notag \\
&&=\int\langle \vec{R}_{\rm cm},\vec{\rho}_{23},\vec{\rho}_{1}|\vec{r}_1,\vec{r}_2,\vec{r}_3\rangle\langle\vec{r}_1,\vec{r}_2,\vec{r}_3|\Psi_{3,0}\rangle d^3\vec{r}_1d^3\vec{r}_2d^3\vec{r}_3 \notag \\
&&=\int \delta\left(\vec{R}_{\rm cm}-\frac{\vec{r}_1+\vec{r}_2+\vec{r}_3}{3}\right)\delta(\vec{\rho}_{23}-\vec{r}_2+\vec{r}_3) \notag \\
&&\times \delta(\vec{\rho}_1-\vec{r}_1+\frac{\vec{r}_2+\vec{r}_3}{2})\phi_0(\vec{r}_1)\phi_0(\vec{r}_2)\phi_0(\vec{r}_3)d^3\vec{r}_1d^3\vec{r}_2d^3\vec{r}_3, \notag \\
\end{eqnarray}
where $\phi_0$ denotes the single-particle ground-state wavefunction of the trap.
By substituting Eq. (\ref{psitran}) into Eq. (\ref{pqlim}) and integrating over $\vec{R}_{\rm cm}, \vec{\rho}_{23}$ and $\vec{\rho}_1$, we obtain
\begin{eqnarray} \label{p0lim}
\lim_{R_0 \to 0} \mathcal{D}_0&=&\int|\phi_0(\vec{r}_1)\phi_0(\vec{r}_2)\phi_0(\vec{r}_3)|^2 \\ \notag
&\times&\delta(\vec{r}_2-\vec{r}_3)\delta(\vec{r}_3-\vec{r}_1)d^3\vec{r}_1d^3\vec{r}_2d^3\vec{r}_3 \\ \notag
&=&\int|\phi_0(\vec{r})|^6d^3\vec{r}.
\end{eqnarray}
As a result, our LL relation (9), when applied to the ground trapped state, reduces to the commonly used \textit{near-threshold} relation: $\Gamma/\hbar = L_3\int d\vec{r}|\phi_{0}(r)|^{6}/3$ \cite{Reynolds:2020,Jack:2002,Ratzel:2021,Goban:2018}. 

We observe that the right-hand side of Eq.~(\ref{p0lim}) can be identified as the three-body correlation function for three particles \cite{Reynolds:2020}. Accordingly, the reactive three-body probability density $\mathcal{D}_q$ defined in the present work can be interpreted as a generalized three-body correlation function for the $q$th trapped three-body state. Equation~(\ref{psr2}), as discussed previously, predicts a linear dependence of the generalized three-body correlation function on the trap frequency $\omega_{\rm ho}$ and a quadratic dependence on the trapped-state energy $E_q$.

\subsection{Time-delay discussion}

We fit the time delay $\tau_D$ to the following Lorentzian form
\begin{equation}
\tau_D(E)=\frac{\hbar\Gamma_q}{(E-E_{q})^2+(\Gamma_q/2)^2}+\tau_{\rm Dbg},
\end{equation}
where an additional parameter $\tau_{\rm Dbg}$ is included to describe the local background time delay. The background parameter $\tau_{\rm Dbg}$ improves the fit performance. We note that with $\tau_{\rm Dbg}$ the width of trapped state is given by $\Gamma_q=4\hbar/[\tau_D(E_q)-\tau_{\rm Dbg}]$, slightly modified from the relation, $\Gamma_q=4\hbar/\tau_D(E_q)$, given by the standard Lorentzian. The background time delay can also provide valuable physical insights into the system. For instance, the sign of its value at the threshold is negative and positive for the $^{85}$Rb and $^{87}$Rb, respectively. This reflects correctly the attractive interaction ($a<0$) between $^{85}$Rb atoms and the repulsive ($a>0$) interaction between $^{87}$Rb atoms at low energies. Nevertheless, further discussions on $\tau_{\rm Dbg}$ are beyond the scope of the present work.

\subsection{Lellouch-L\"uscher relation for $N$-body losses}
\begin{figure}[t]
 \centering
  \resizebox{0.45\textwidth}{!}{\includegraphics{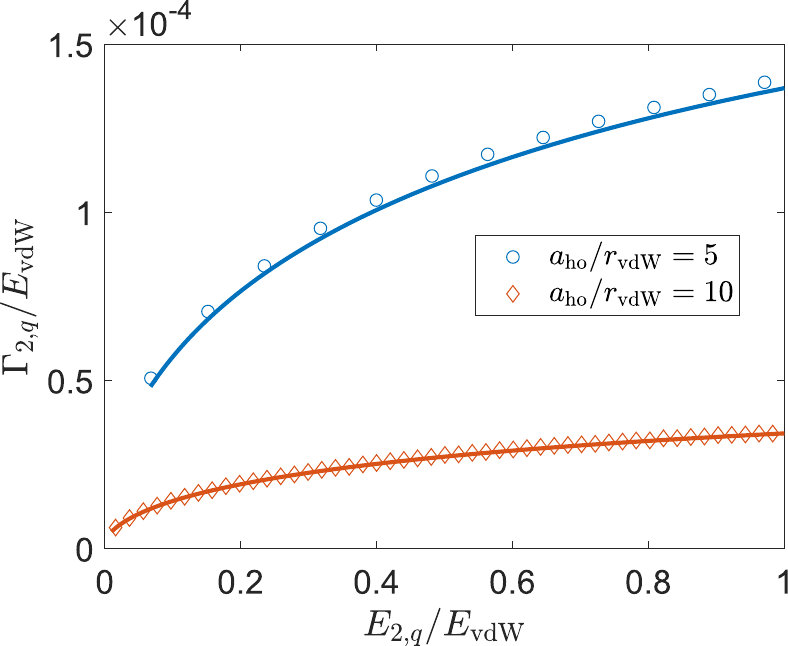} }
 \caption{\label{fig:ll2b} The width $\Gamma_{2,q}$ of trapped two-body state versus its energy $E_{2,q}$ for $^{87}$Rb in $|2,-1\rangle$ state at $B=1000$ G. The data symbols represent results obtained by fitting the two-body time delay at $a_{\rm ho}/r_{\rm vdW} =5$ (circles) and 10 (diamonds) to a Lorentzian form, while the solid line corresponds to the prediction from Eq. (\ref{eq:ll2b}).}
\end{figure}
In this section, we generalize the LL relation (9) to connect the free-space $N$-body loss rate coefficient $L_N (E_{N,q})$ \cite{explainLN,mehta2009prl} with the energy $E_{N,q}$ and width $\Gamma_{N,q}$ of a trapped $N$-body state, and the trap frequency $\omega_{\rm ho}$. We initially consider $N$-body loss (i.e. each loss event involving $N$ particles) as the dominant decay pathway of an $N$-body state, but will later show that this limitation is not essential. Consequently, we have
\begin{equation} \label{2bGamma1}
\frac{\Gamma_{N,q}}{\hbar}=\frac{L_N(E_{N,q})}{N}\mathcal{D}_{N,q},
\end{equation} 
where the $N$-body reactive probability density $\mathcal{D}_{N,q}$ is given by
\begin{eqnarray} \label{pqr}
\mathcal{D}_{N,q}=\frac{1}{\mathcal{V}_{N,0}}\int_0^{R_{N,0}}F_{N,q}(R_N)^2dR_N,
\end{eqnarray}
within the hyperspherical volume $\mathcal{V}_{N,0}=\pi^{\frac{3N-3}{2}}R_{N,0}^{3N-3}/\Gamma(\frac{3N-1}{2})$ \cite{sphv} of the reaction regime ($R_N<R_{N,0}$). Here, $R_{N}$ is the $N$-body hyperradius describing the overall size and $\Gamma(\cdot)$ denotes the Gamma function. The non-interacting $N$-body wavefunction reads 
\begin{eqnarray}
F_{N,q}(R_N)&=&\frac{1}{\sqrt{M_{N,q}}}R_{N}^{\frac{3N-4}{2}}e^{[-R_N^2/(2 N^{\frac{1}{N-1}}a_{\rm ho}^2)]} \\ \notag
&\times& L_q^{(\frac{3N-5}{2})}[R_N^2/(N^{\frac{1}{N-1}} a_{\rm ho}^2)],
\end{eqnarray}
with a normalization factor $M_{N,q}$ given by
\begin{equation}
M_{N,q}=\frac{N^{3/2}\Gamma(q+\frac{3N-3}{2})}{2\Gamma(q+1)}a_{\rm ho}^{3N-3}.
\end{equation} 
Similar to the previous derivation in the three-body case, we assume $R_{N,0}\ll a_{\rm ho}$ and retain only the leading-order contribution:
\begin{eqnarray}
\mathcal{D}_{N,q}&=&\frac{1}{M_{N,q}\mathcal{V}_{N,0}}\left(\frac{\Gamma(q+\frac{3N-3}{2})}{q!\Gamma(\frac{3N-3}{2})}\right)^2\int_{0}^{R_{N,0}}R_N^{3N-4}dR_N \notag \\
&=&\frac{1}{a_{\rm ho}^{3N-3}\pi^{\frac{3N-3}{2}}N^{3/2}\Gamma(\frac{3N-3}{2})} \frac{\Gamma(\frac{3N-3}{2}+q)}{\Gamma(1+q)}, \label{DNq}
\end{eqnarray}
or, defining $q$ from its relationship with $E_{N,q}=[2q+(3N-3)/2]\hbar \omega_{\rm ho}$, 
\begin{align}
\mathcal{D}_{N,q}
&=\frac{\Gamma(\frac{E_q}{2\hbar\omega_{\rm ho}}+\frac{3N-3}{4})/\Gamma(\frac{E_q}{2\hbar\omega_{\rm ho}}-\frac{3N-7}{4})}{a_{\rm ho}^{3N-3}\pi^{\frac{3N-3}{2}}N^{3/2}\Gamma(\frac{3N-3}{2})}, \label{DnqNew}
\end{align}
The above equation can be further simplified by employing the asymptotic form of the function below for large $x$:
\begin{align}
    \frac{\Gamma(x+\frac{3N-3}{4})}{\Gamma(x-\frac{3N-7}{4})}
    =x^{\frac{3N-5}{2}}\left(1-\frac{\alpha_{N}}{x^2}+\frac{\beta_{N}}{x^4}+\cdots\right)\label{GammaX}
\end{align}
where $\alpha_N$=$(N$-$1)(3N$-$7)(3N$-$5)/64$, and $\beta_N$=$(N$-$3)(N$-$1)(3N$-$11)(3N$-$7)(3N$-$5)(15N$-$11)/40960$. For simplicity, we truncate to the $1/x^2$ term and get 
\begin{align}
\mathcal{D}_{N,q}
&\simeq\frac{(\frac{E_q}{2\hbar\omega_{\rm ho}})^{\frac{3N-5}{2}}[1-\alpha_N/(\frac{E_q}{2\hbar\omega_{\rm ho}})^2]}{a_{\rm ho}^{3N-3}\pi^{\frac{3N-3}{2}}N^{3/2}\Gamma(\frac{3N-3}{2})}
\nonumber\\
&=\frac{(\frac{1}{2\hbar\omega_{\rm ho}})^{\frac{3N-5}{2}}E_q^{\frac{3N-9}{2}}[E_q^2-4\alpha_N (\hbar\omega_{\rm ho})^2]}{a_{\rm ho}^{3N-3}\pi^{\frac{3N-3}{2}}N^{3/2}\Gamma(\frac{3N-3}{2})}.
\label{DnqEq}
\end{align}
Note that for $N=3$ the above expression is exact because $\beta_3=0$ and all higher-order coefficients vanish.
For other values of $N$, the asymptotic expression for $\mathcal{D}_q$ above leads to a relative error below 10\% for $E_{q=0}$ and $N\leq5$, and decreases rapidly as $1/E_q^4$ for larger values of $E_{q>0}$ for all $N$.
By substituting Eq. (\ref{DnqEq}) into Eq. (\ref{2bGamma1}), we then get
\begin{align} 
\Gamma_{N,q}
&=\frac{1}{2^{\frac{3N-5}{2}}\pi^{\frac{3N-3}{2}}N^{\frac{5}{2}}\Gamma(\frac{3N-3}{2})}\frac{m^{\frac{3N-3}{2}}}{\hbar^{3N-5}}L_N(E_{N,q}) \nonumber\\
&\times E_{N,q}^{\frac{3N-9}{2}}[E_q^2-4\alpha_N (\hbar\omega_{\rm ho})^2]\omega_{\rm ho}, \label{eq:ll2b}
\end{align}
leading to the result expressed in Eq.~(10) of the main text. In the case 
$N=3$, the above expression subsumes Eq.~(9), with its validity being demonstrated in the main text. For $N=2$, we also demonstrate the validity of Eq. (\ref{eq:ll2b}) by considering $^{87}$Rb atoms in $|2,-1\rangle$ state and calculating free-space two-body loss rate as well as the width and energy of the trapped two-body state at $B=1000$ G, as is shown in Fig. \ref{fig:ll2b}. Here, the decay originates from the relaxation of atoms from higher to lower hyperfine states. A magnetic field of
$B=1000$ G is chosen to guarantee that the released energy is sufficient for the atoms to leave the trap following decay.

%\bibliography{biblio}

\begin{thebibliography}{87}
\bibitem [{exL()}]{exL3}%
  \BibitemOpen
  \href@noop {} {}\bibinfo {note} {{We note that while $\Gamma_{q}$ is the exponential decay rate of an single quantum state, whereas in a bulk gas $L_3$ controls the rate of successive decays events, depleting the atom number and resulting in a non-exponential decay for the atomic density, $n$, according to: $dn/dt=-L_3n^3$.}}\BibitemShut {Stop}%
\bibitem [{exp()}]{explainLN}%
  \BibitemOpen
  \href@noop {} {}\bibinfo {note} {In ultracold quantum gases, the $N$-body loss rate coefficient, $L_N$, is defined such that the time evolution of atomic density, $n$, is given by: $dn/dt=-\sum_{N=2}L_Nn^N$.}\BibitemShut {Stop}%
\end{thebibliography}
%apsrev4-2.bst 2019-01-14 (MD) hand-edited version of apsrev4-1.bst
%Control: key (0)
%Control: author (8) initials jnrlst
%Control: editor formatted (1) identically to author
%Control: production of article title (0) allowed
%Control: page (0) single
%Control: year (1) truncated
%Control: production of eprint (0) enabled
%

\end{document}